# Opportunistic Lower-Terahertz Rainfall Estimation with DSD-Constrained Channel Characterization

Wanzhu Chang, Yuheng Song, Jiabiao Zhao, Mingxia Zhang, Jie Yang, Kefeng Huang, Kaixin Sun, Hong Liang, Weidong Hu, Fawad Sheikh, *Member, IEEE*, and Jianjun Ma, *Member, IEEE*

***Abstract*—Rain-induced attenuation and scattering become significant at terahertz (THz) frequencies, and exploiting this rain sensitivity is necessary both to safeguard link reliability and to enable opportunistic environmental sensing without dedicated instrumentation, a capability that remains largely unvalidated on real outdoor channels above 100 GHz. This article investigates opportunistic rainfall estimation using measured lower-terahertz (THz) channels at 140 and 229 GHz. Outdoor measurements over a 41.5-m rain-exposed path are used to characterize rain-induced attenuation and the rainfall dependence of an effective Rician K-factor. Because the path-representative drop-size distribution (DSD) is unavailable, several propagation-model scenarios based on ITU-R P.838-3 and Mie theory with canonical DSDs are employed to quantify model-form sensitivity. These channel characteristics are then used to generate physics-constrained synthetic received-power sequences for training RainFormer, a compact attention-convolution regression network that combines temporal features with explicit attenuation and fluctuation statistics. Under matched synthetic conditions, RainFormer achieves RMSEs of 0.1782 and 0.2925 mm/h at 140 and 229 GHz, respectively, and outperforms the investigated convolutional and Transformer baselines in most metric-frequency combinations. Direct application to the independent measured dataset produces physically consistent rainfall estimates at 140 GHz and demonstrates that received-power fluctuations provide useful information beyond mean attenuation. The results establish a measurement-informed framework for evaluating lower-THz links as opportunistic rainfall sensors while explicitly accounting for propagation-model uncertainty.**

## I. INTRODUCTION

Wireless networks are expected to integrate high-capacity communication with native environmental sensing, making integrated sensing and communication (ISAC) an important paradigm for 6G and beyond [1],[2]. The terahertz (THz) band has attracted increasing attention because its ultra-wide bandwidth, short wavelength, and highly directional beams can support both high-throughput links and fine-resolution sensing [3],[4]. However, the THz channel is strongly affected by atmospheric conditions, particularly molecular absorption and weather-induced extinction. Rain and snow cause pronounced attenuation at THz frequencies [5]-[9], reaching about 20 dB/km at 50 mm/h rainfall rate [10]. This loss depends on carrier frequency, path length, temperature, polarization, and raindrop size distribution (DSD) [11], with DSD variability and propagation geometry strongly influencing modeling accuracy [12],[13]. These perturbations are both a reliability challenge and a physically interpretable source of meteorological information. Received-power variations can map to rainfall intensity, letting a THz channel act as an opportunistic rainfall sensor while retaining its communication function [3].

Mature operational rainfall products rely on dedicated meteorological sensors, long-term calibration, and large-scale data fusion. The Global Precipitation Measurement (GPM) mission [14] retrieves precipitation structure and microphysics by combining dual-frequency precipitation radar with a 13-channel microwave imager spanning 10.65-183.31 GHz. Derived products such as IMERG [15], CMORPH [16], and GSMaP [17] further merge multi-satellite passive-microwave retrievals (~10-190 GHz) with infrared information to produce gridded global precipitation fields. Notably, these systems exploit channels extending above 100 GHz, where the scattering and emission signatures of small hydrometeors become strong enough to detect.

Beyond these products, deep learning has been widely applied to satellite- and radar-based rainfall estimation and nowcasting, including convolutional [18],[19], generative-adversarial [20], and U-Net/ConvLSTM architectures [21]-[23] that map multisource microwave and infrared observations or radar composites to precipitation intensity and extend nowcast lead times. In parallel, opportunistic sensing with commercial microwave links (CMLs) has been studied extensively in the microwave and millimeter-wave bands. Received-signal levels from cellular networks were shown convertible to surface rainfall and validated against radar and gauges [24],[26], and learning-based processing followed, including CNN-based rain-event detection on large operational link sets [27], alongside reviews covering the standard wet-dry classification, baseline, wet-antenna, and attenuation-to-rain-rate processing chain in the 5-40 GHz range [28],[29]. All of these depend on large-scale satellite, radar, or dense operational microwave-link datasets already available for supervised learning. By contrast, THz-channel-based rainfall estimation lacks comparable operational datasets, with existing THz rain-channel studies limited to controlled artificial-rain experiments or short-term outdoor campaigns [7]-[12].

To address the scarcity of labeled outdoor THz rainfall data and the uncertainty introduced by the unknown path-representative DSD, this article combines measured channel characterization with physics-constrained synthetic training. The resulting framework uses both attenuation and short-term received-power fluctuations to estimate rainfall while explicitly evaluating sensitivity to alternative propagation models. Section II presents the measurement campaign and channel characterization, Section III describes the synthetic-data generation procedure, Section IV introduces RainFormer, Section V discusses the estimation results, and Section VI concludes the article.

## II. Channel measurement and characterization

The measurement data used in this work were acquired during an outdoor rainfall experiment conducted at Beijing Institute of Technology (Liangxiang campus), in August 2025. The experimental setup used a fixed point-to-point THz channel, as illustrated in Fig. 1(a). At the transmitter, a Ceyear 1465D signal generator provided the source signal over 100 kHz-20 GHz. For the 110-170 GHz band, a Ceyear 82406B frequency multiplier with a multiplication factor of ×12 was used. For the 220-325 GHz band, a Ceyear 82406D frequency multiplier with a multiplication factor of ×18 was used. The resulting continuous-wave THz signal was radiated by a Ceyear 89901S horn antenna and collimated by a high-density polyethylene lens with a focal length of 30 cm. At the receiver, the same horn-lens configuration collected the signal, which was directly detected by a Ceyear 71718 power sensor. Both the transmitter and the receiver were mounted 122 cm above the ground. This height provided sufficient clearance relative to the first Fresnel-zone radius, which was approximately 14.9 cm at 140 GHz and 11.7 cm at 229 GHz. The rain-exposed channel length was 41.5 m. We recorded the channel response under both clear and rainy conditions, with their difference used to extract the rain-induced attenuation.

Both transmitter and receiver use only vertical linear polarization throughout the measurement campaign, due to the fixed orientation of the horn-lens optics and the limited duration of the available rainfall event. Signal generation and power acquisition were coordinated by a host computer over a LAN/RS-485 instrument-control network (LAN-SCPI commands), continuously recording the temporal power variations introduced by rainfall. During the campaign the horn-lens assemblies at both ends were sheltered from direct wetting, so the extracted attenuation is dominated by path rain rather than by antenna or lens wetting. This removes the wet-antenna term that is a significant and widely studied error source in commercial-microwave-link rainfall estimation, alongside coarse signal quantization, wet/dry classification, and baseline determination.

We repeated single-frequency acquisitions as rainfall intensity evolved, with each acquisition limited to less than 1 min to reduce rain-rate nonstationarity within the observation interval and the temporal mismatch between the received-power record and the rain-gauge measurement [30]. Ground truth was provided by a point rain gauge located near the center of the rain-exposed path. The received-power records were segmented into non-overlapping 20 s windows according to their recorded timestamps. Windows with fewer than 30 valid observations were excluded. To maintain dimensional compatibility with the fixed 400-position input of the pretrained RainFormer, the observations in each retained window were mapped from their acquisition-order indices onto a normalized 400-position grid using piecewise cubic Hermite interpolation. Resampling was used solely to construct a fixed-dimensional network input; it was not assumed to recover unobserved high-cadence channel fluctuations, create new physical measurements, or increase the effective sample size. Each 20 s window was treated as one statistical observation, and the 400 positions within the resulting input vector were regarded as temporally correlated representations of that window. After quality control, the resulting *BIT-25* dataset contained 29 window-level samples at 140 GHz and 6 at 229 GHz. The window-level mean attenuation was used only for the attenuation – rainfall comparison in Fig. 1(b), whereas the resampled and normalized sequence was used as the input to the measured-data consistency assessment. All prediction errors and uncertainty estimates were calculated at the window level.

Fig. 1(b) compares the measured attenuation with DSD-prior-based attenuation curves at 140 GHz and 229 GHz, computed using the ITU-R P.838-3 model [31] and Mie theory with several empirical DSD models, as Marshall-Palmer (M-P) [32], Joss-Widespread (J-W) [33], and Joss-Drizzle (J-D) [34]. More details about the theoretical models can be found in reference [6]. At 140 GHz all samples lie in the weak-rain regime near the J-D, M-P, and J-W curves. At 229 GHz the samples correspond to higher rain rates and larger attenuation, lying closer to the upper J-D curves. The Joss-Thunderstorm prior [34] was excluded because, at both frequencies, it produced an attenuation-rainfall relationship well below the measured envelope and the other four priors, adding configuration count without additional bracketing. The four propagation-model scenarios (ITU-R P.838-3, M-P, J-W, J-D) span the physically relevant range of microphysical assumptions. Because the channel representative DSD cannot be directly observed with conventional point disdrometers - a recognized open problem in operational radar meteorology [35],[36] - we bracket the unknown DSD across these canonical priors rather than asserting a single model, and report the resulting spread as a model-form spread.

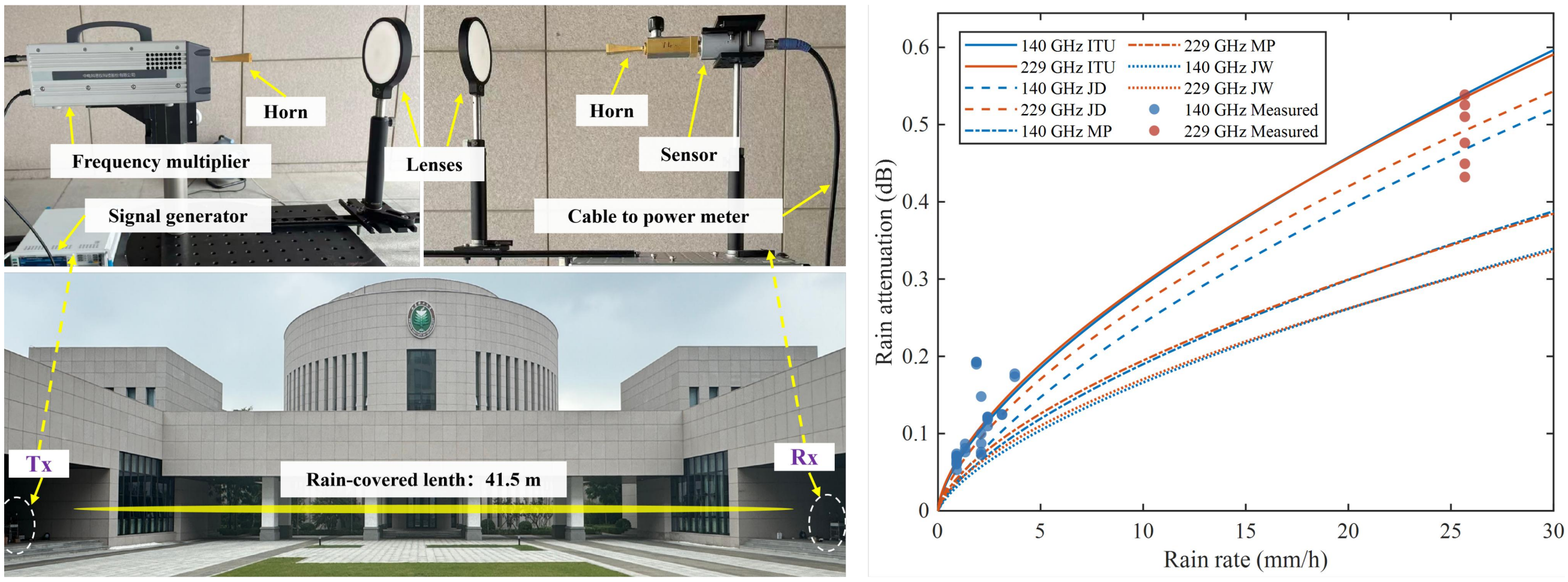

(a) (b)

**Fig. 1.** Outdoor THz rainfall measurement setup and measured attenuation sample distribution. (a) Fixed point-to-point THz channel with a 41.5 m rain-exposed path. (b) Measured rain-induced attenuation samples and DSD-prior-based attenuation curves at 140 and 229 GHz.

Ambient conditions were stable throughout the operational window ($T$ = 27 ºC, RH = 71.5%), with clear-sky and rainy measurements acquired in a single continuous instrument-on period on August 4, 2025. A worst-case ITU-R P.676-13 bound, $\Delta A_{\text{atm}} \leqslant L\cdot(|\partial\gamma/\partial\rho_w|\ \Delta\rho_w + |\partial\gamma/\partial T|\ \Delta T)$, with credible bounds of $\Delta\rho_w \leqslant 0.5$ g/m$^3$ and $\Delta T \leqslant 0.5$ ºC over the 41.5 m path, gives a baseline atmospheric drift below 0.005 dB at 140 GHz and 0.008 dB at 229 GHz. They are under the ±0.043 dB (±1.0%) power-meter uncertainty and far below the ITU-R P.838-3 rain attenuation expected over the same path. The rain-induced attenuation is therefore dominated by rain extinction rather than atmospheric or instrumental baseline drift.

It should be noted that the weakest measured rain-induced attenuation, approximately 0.06 dB, is comparable to the power-meter uncertainty. Nevertheless, rainfall information is not represented solely by the window-mean attenuation. Rainfall also modulates the short-term received-power fluctuation statistics, including the effective Rician K-factor, allowing the estimator to exploit complementary information from both attenuation level and temporal variation.

To characterize the small-scale fading statistics under rainfall, we relate the Rician $K$-factor to the rainfall rate $R$. Prior work shows that in rain-attenuation scenarios the $K$-factor decreases approximately linearly with $R$ [37], as

$$K_{dB}(f,R)=K_0(f)-a(f)R \tag{1}$$

where $K_0(f)$ is the extrapolated zero-rain intercept (baseline $K$-level near the weak-rain limit) and $a(f)$ is the rainfall sensitivity coefficient. To avoid information leakage into the measured-data transfer evaluation, the $K$-factor fitting uses an independent dataset, *BIT-24*, acquired at Beijing Institute of Technology in August 2024 and reported in reference [6], rather than the *BIT-25* in this work. Ordinary least-squares fitting yielded slopes $a(f)$ of 0.2150 and 0.2765 and intercepts $K_0(f)$ of 50.483 and 50.311 dB at 140 GHz and 229 GHz, respectively. 5000 bootstrap resampling trials [38] gave 5$^{\text{th}}$-95$^{\text{th}}$ percentile intervals of (48.677, 52.073) at 140 GHz and (48.436, 51.913) at 229 GHz. The $K$-factor decreases with rising rainfall rate at both frequencies. This behaviour reflects the progressive erosion of the dominant line-of-sight component as attenuation accumulates along the rain-exposed span, rather than a claim of rain-induced diffuse scattering. The fitted intercepts correspond to linear-scale $K$-factors of order $10^5$, consistent with the directive horn-lens optics and the short unobstructed path, so the Rician term acts as a bounded stochastic perturbation rather than a multipath model.

The linear $K$-$R$ parameterization is applied within the rain-rate range spanned by the *BIT-24* fit, and the coefficients $a(f)$ are used only at the two measured carriers rather than interpolated in frequency. Extrapolation beyond either range is not implied.

## III. Synthetic Data Generation

Using the $K$-$R$ relation, we construct a physics-constrained synthetic dataset for continuous rainfall-rate regression that jointly models large-scale rain attenuation, small-scale Rician fading, and slow background channel drift. For a given frequency $f$ and rainfall rate $R$, the specific rain attenuation $\gamma_{rain}(R, f, T)$ is obtained from the vertical-polarization DSD priors of Section II, multiplied by the channel length to give the total path attenuation, which sets the mean received-power level of the generated sequence. All channel parameters match the measured configuration - length $L$ = 41.5 m, temperature $T$ = 27 ºC, and vertical polarization.

Small-scale fluctuations follow a time-varying Rician model in which the fitted $K$-$R$ relation supplies the rainfall-dependent baseline $K$-factor, and a bounded first-order autoregressive perturbation emulates slow channel-statistical drift within the 20 s window. This drift is motivated by measured nonstationary channels with a time- and frequency-varying K-factor [39], and by THz rainfall measurements showing rain-induced loss and DSD statistics varying dynamically even at similar rain rates [6, 12]. After converting the $K$-factor to linear scale, the complex channel coefficient

can be generated as

$$h[k]=\sqrt{\frac{K[k]}{K[k]+1}}e^{j\phi_0}+\sqrt{\frac{1}{K[k]+1}}g[k] \tag{2}$$

where $\phi_0$ is the fixed direct-path phase within the window and $g[k]$ is the scattered component with unit average power, following the standard Rician formulation in which $K$ is the deterministic-to-diffuse power ratio [40].

The resulting fading sequence is applied around the window-level mean power to produce a 400-point received-power sequence at a 0.05 s sampling interval, so each sample is a 20 s sequence paired with one constant rainfall-rate label. Because rain events are long-tailed, with light rain far outnumbering heavy rain [41], labels are drawn from a two-component uniform mixture - $R \in [1,10]$ mm/h with probability 0.7 and $R \in [10,50]$ mm/h with probability 0.3. For each frequency and DSD prior, 20,000 samples are generated. The resulting dataset, denoted *BIT-24-Gen*, built from the leakage-free *BIT-24 K-R* parameters of Section II, forms the training basis for rainfall-rate regression, while the measured *BIT-25* set is reserved for the transfer evaluation.

## IV. Hybrid attention convolution regression network

RainFormer maps a short-time THz received-power sequence to a scalar rainfall rate. The input is a normalized single-channel sequence of length 400 (a 20 s window at 0.05 s sampling, as defined in Section III), and the output is the rainfall-rate estimate for that window. The design is governed by a single observation about the signal - under rainfall, the received power superimposes a slow attenuation trend on fast small-scale fluctuations, so an effective estimator must model both long-range temporal structure and local variation. As shown in Fig. 2, RainFormer realizes this through a one-dimensional convolutional embedding, four stacked hybrid attention-convolution blocks operating in a 128-dimensional latent space, attentive temporal pooling, and a regression head with explicit physical-statistical fusion.

The convolutional embedding applies a depthwise temporal convolution for local preprocessing and projects the sequence into the latent space. A sinusoidal positional encoding then preserves temporal order before the backbone. Each hybrid block couples global and local modeling. A multi-head self-attention sublayer captures correlations across the full window, and depthwise-convolutional gated feed-forward sublayer refines local nonlinear fluctuation patterns. For the $l$-th block with input $Z^{(l-1)}$,

$$\tilde{Z}^{(l)}=Z^{l-1}+\mathrm{Drop}(\mathrm{MHA}(\mathrm{LN}(Z^{l-1}))) \tag{3}$$

$$Z^{(l)}=\tilde{Z}^{(l)}+\mathrm{Drop}(\mathbf{W}_2[\mathrm{GLU}(\mathbf{W}_1(\mathrm{DWConv}(\mathrm{LN}(\tilde{Z}^{(l)}))))]) \tag{4}$$

where $\mathrm{LN}(\cdot)$ is layer normalization [42], $\mathrm{MHA}(\cdot)$ is multi-head self-attention for long-range temporal dependencies [43], $\mathbf{W}_1$ and $\mathbf{W}_2$ are learnable projections, $\mathrm{GLU}(\cdot)$ is a gated linear unit [44], and $\mathrm{Drop}(\cdot)$ is dropout [45]. Each block thus jointly represents the long-range dependencies and local fluctuation structure that characterize rainfall-modulated received power. The backbone is complemented by an explicit physical statistical pathway. The mean and standard deviation of the original normalized input (first- and second-order descriptors of the window's attenuation level and fluctuation magnitude) are concatenated with the attentively pooled latent representation and passed to a multilayer regression head. Such fusion ensures that the dominant attenuation and fluctuation statistics, which carry direct physical information about rain intensity, are retained explicitly rather than relying on the backbone to recover them from the sequence alone [46].

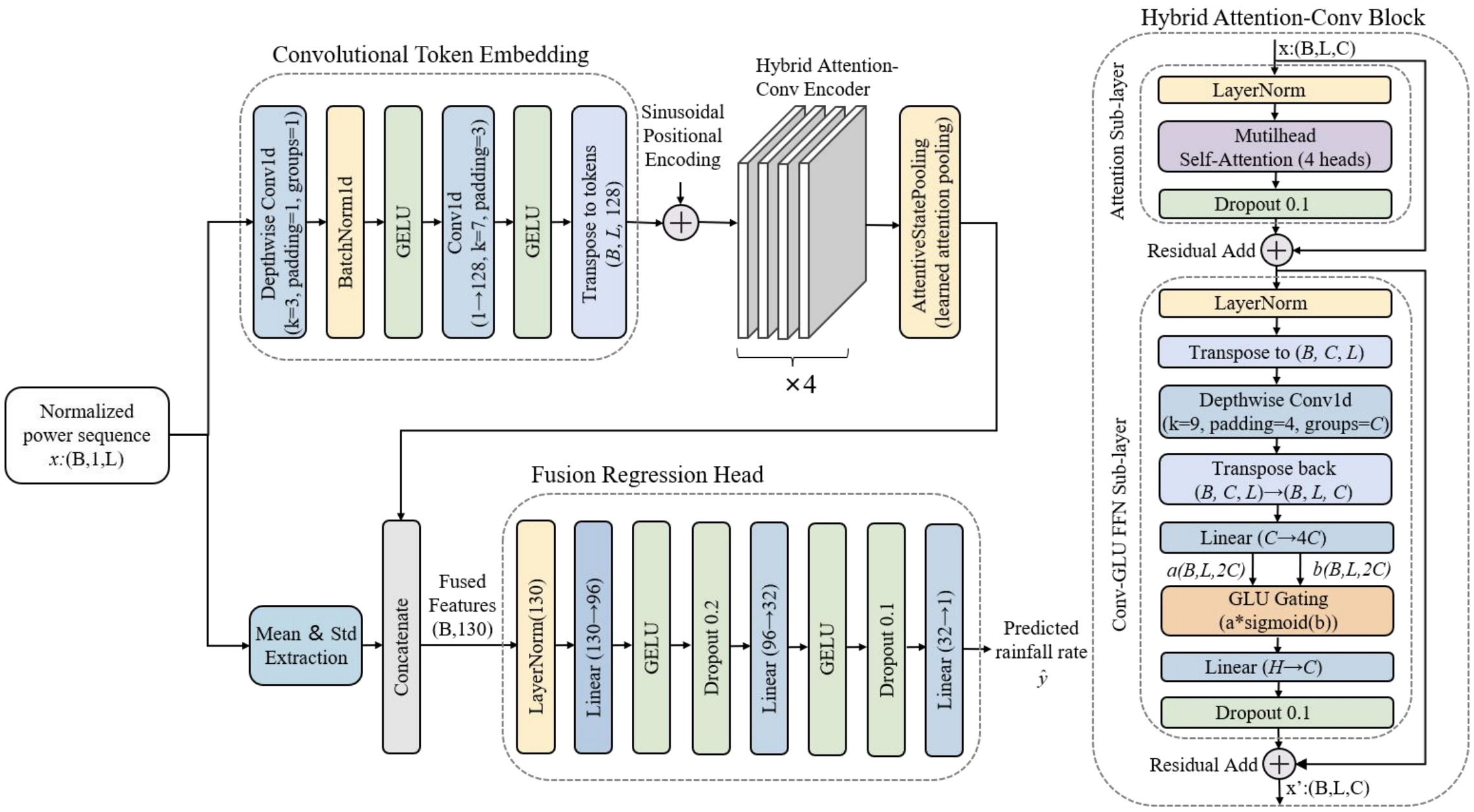

**Fig. 2.** Architecture of the proposed RainFormer network for THz received-power sequence regression. The left portion illustrates the main network backbone, whereas the right portion details the internal structure of the hybrid attention-convolution block.

TABLE I

KEY TRAINING HYPERPARAMETERS OF RAINFORMER

| Hyperparameter | Value |
|---|---|
| Optimizer | Adam |
| Loss function | MSE |
| Initial learning rate | $1\times10^{-3}$ |
| Batch size | 64 |
| Maximum training epochs | 30 |
| LR scheduler patience (epochs) | 3 |
| LR decay factor | 0.5 |

RainFormer is trained independently for each frequency-DSD-prior configuration on the corresponding *BIT-24-Gen* subset. The 20 thousand samples per configuration are randomly partitioned into training, validation, and test sets in a 70/15/15 ratio, with a fixed random seed ensuring reproducible partitioning and initialization. The network is optimized as a supervised regressor by minimizing the mean squared error between predicted and reference rainfall rates, using Adam with an initial learning rate of $1\times10^{-3}$ and batch size 64. The learning rate is halved after three consecutive epochs without validation improvement. Training runs for at most 30 epochs with early stopping (patience 8), and the checkpoint with the lowest validation loss is retained for testing. The hyperparameters are summarized in Table I.

## V. RESULTS AND DISCUSSION

### *A. In-distribution performance on synthetic data*

We first assess RainFormer under matched conditions, training and testing on *BIT-24-Gen* to verify that the network learns the intended mapping from received-power sequences to rainfall rate. Table II reports in-distribution RMSE and MAE at 140 and 229 GHz across the four DSD priors. The optimal prior is frequency-dependent. At 140 GHz the ITU prior gives the lowest error (RMSE 0.1782, MAE 0.1107 mm/h), with the empirical priors moderately higher. At 229 GHz the M-P prior is best (RMSE 0.2925, MAE 0.1916 mm/h) and ITU-R no longer leads. That the best prior changes with frequency shows no single DSD assumption is uniformly optimal - regression difficulty is set jointly by the frequency dependence of rain attenuation and the DSD-dependent statistical structure of the generated fluctuations.

Fig. 3 examines these differences through error distributions as Fig. 3 (a) and (d), and conditional-density plots as Fig. 3 (b), (c), (e) and (f). At 140 GHz the ITU-R prior yields a tighter zero-centered error distribution than M-P, consistent with Table II. At 229 GHz this reverses, M-P producing the narrower distribution. In all conditional-density panels the high-density ridge tracks the 1:1 reference and the conditional mean $E(\hat{R}\,|\,R)$ follows the diagonal, indicating that under matched conditions RainFormer learns a stable, near-monotonic mapping. This confirms the internal consistency and learnability of the physics-constrained datasets - the in-distribution validation step that sim-to-real pipelines require before transfer can be meaningfully assessed [47], [48]. Whether the learned mapping survives the synthetic-to-measured domain gap is evaluated in Section V-D.

TABLE II

PERFORMANCE BENCHMARK ON *BIT-24-GEN* UNDER VARIOUS

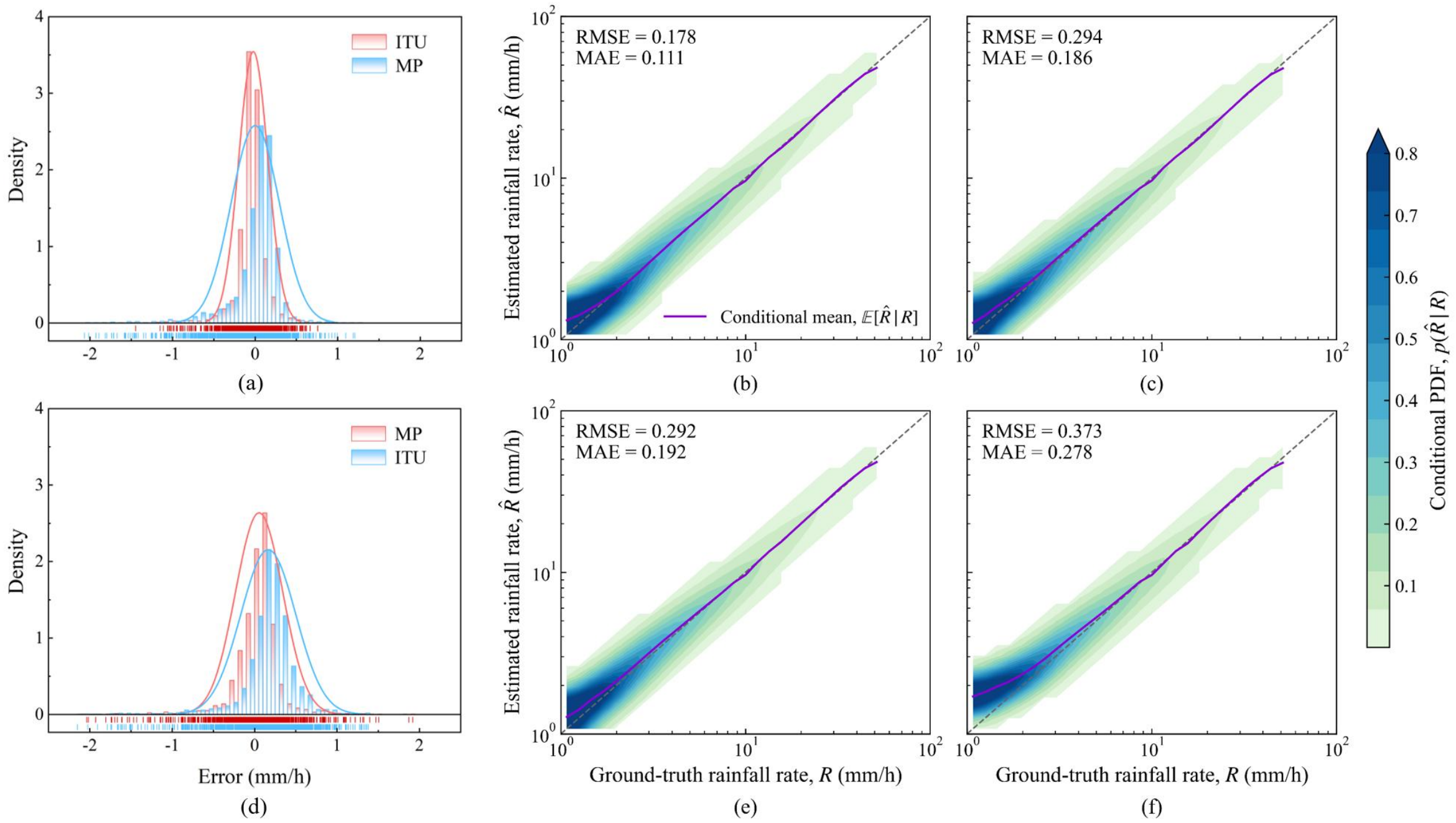

**Fig. 3.** Error-distribution and conditional-density analysis of RainFormer on matched synthetic datasets. Panels (a) and (d) compare the estimation-error distributions, kernel-density curves, and rug marks of the best- and worst-performing DSD priors at 140 and 229 GHz, respectively. Panels (b) and (c) correspond to the two 140 GHz cases in panel (a), namely the ITU and M-P priors, while panels (e) and (f) correspond to the two 229 GHz cases in panel (d), namely the M-P and ITU priors. In the conditional-density plots, the gray dashed line denotes the ideal 1:1 relationship.

THEORETICAL DSD ASSUMPTIONS, RMSE AND MAE ARE REPORTED IN MM/H.

| DSD model | RMSE | | MAE | |
|---|---|---|---|---|
| | 140GHz | 229GHz | 140GHz | 229GHz |
| ITU | **0.1782** | 0.3735 | **0.1107** | 0.2784 |
| JD | 0.2447 | 0.3587 | 0.1570 | 0.2318 |
| MP | 0.2938 | **0.2925** | 0.1859 | **0.1916** |
| JW | 0.2864 | 0.3522 | 0.1853 | 0.2188 |

TABLE III
ABLATION STUDY OF RAINFORMER COMPONENTS UNDER THE BEST DSD-PRIOR CONFIGURATION AT EACH FREQUENCY.

| Model | RMSE | | MAE | |
|---|---|---|---|---|
| | 140GHz | 229GHz | 140GHz | 229GHz |
| RainFormer | 0.1782 | 0.2925 | 0.1107 | 0.1916 |
| w/o Physical Statistics | 0.7841 | 1.6515 | 0.4196 | 0.8994 |
| Mean Pooling | 0.6665 | 0.3018 | 0.4857 | 0.1844 |
| w/o Positional Encoding | 1.1770 | 1.5891 | 0.9302 | 0.9269 |
| w/o Attention | 0.2080 | 0.3206 | 0.1372 | 0.2057 |
| w/o Depthwise Conv. | 0.7080 | 0.4985 | 0.4570 | 0.3833 |
| w/o Gated FFN | 0.3578 | 0.3854 | 0.2813 | 0.2399 |
| Plain Patch Embedding | 0.2475 | 0.3545 | 0.2026 | 0.2654 |

### B. Ablation study

To quantify each component's contribution, we ablated RainFormer under the best prior at each frequency (ITU at 140 GHz, M-P at 229 GHz), removing or replacing one component at a time - physical-statistical fusion, attentive pooling, positional encoding, self-attention, depthwise convolution, gated feed-forward, and the convolutional embedding - while holding the *BIT-24-Gen* split and optimization protocol fixed. Table III reports the results.

The ablation results are frequency dependent. At 140 GHz, removing the positional encoding produces the largest RMSE degradation (1.1770), exceeding the physical-statistics ablation (0.7841). The remaining ablations rank below these two at 140/229 GHz. Removing the depthwise convolution raises the RMSE to 0.7080/0.4985, the gated feed-forward network to 0.3578/0.3854, the plain patch embedding to 0.2475/0.3545, and mean pooling to 0.6665/0.3018, versus 0.1782/0.2924 for the full RainFormer. Mean pooling exhibits a pronounced frequency asymmetry, degrading the 140 GHz estimate substantially more than the 229 GHz estimate, consistent with the weaker 140 GHz attenuation signal relying more heavily on attention-weighted temporal aggregation [43],[49]. The overall importance hierarchy is thus positional encoding > physical statistics > depthwise convolution / mean pooling > gated FFN > patch embedding > attention.

### C. Comparison with baseline models

We compare RainFormer against two baselines spanning the dominant sequence-modeling paradigms - a 1D-CNN (local convolutional extraction) and a vanilla Transformer (global attention). The 1D-CNN is a compact, strong end-to-end baseline using shared local filters on the raw sequence. 1D-CNNs have been used to detect rain-induced attenuation in commercial microwave-link signals and to build radar-based QPE [27], [50], and convolutional models have been applied to communication-link precipitation sensing [51]. The vanilla Transformer models pairwise interactions via self-attention, the standard backbone for sequence representation [52], [53], and has been adopted for precipitation modeling [23]. All three use the same split and protocol. Table IV reports the comparison.

At 140 GHz RainFormer achieves the lowest RMSE and MAE (0.1782 and 0.1107 mm/h), reducing RMSE by approximately 13.0% and MAE by 15.3% relative to the 1D-CNN, and by 40.3% and 47.1% relative to the vanilla Transformer. At 229 GHz RainFormer gives the lowest MAE (0.1916 mm/h versus 0.2393 and 0.2066 mm/h), while its RMSE (0.2925 mm/h) is below the 1D-CNN and numerically indistinguishable from the vanilla Transformer (0.2929 mm/h). The joint behavior of the two metrics is informative. Near-equal RMSE with lower MAE implies that RainFormer's accuracy gain is concentrated in typical samples while extreme errors remain comparable, since RMSE weights large deviations quadratically and MAE weights all deviations uniformly [54]. The frequency dependence of the margin is consistent with the ablation in Section V-B. At 140 GHz, where rain attenuation is weaker and the informative signal resides partly in local fluctuation structure, the convolutional pathway and explicit statistical descriptors yield a decisive advantage over pure attention; at 229 GHz, where the attenuation level itself dominates the mapping, global sequence models recover most of the available information and the architectural margin narrows. The hybrid design thus does not uniformly outperform its constituents but degrades gracefully toward the stronger paradigm at each frequency - the principal benefit of convolution-augmented attention architectures [55] - making it the only model among the three that is never the weakest under any metric–frequency combination.

TABLE IV
BASELINE COMPARISON ON *BIT-24-GEN* UNDER THE BEST DSD-PRIOR CONFIGURATION AT EACH FREQUENCY.

| Model | RMSE | | MAE | |
|---|---|---|---|---|
| | 140GHz | 229GHz | 140GHz | 229GHz |
| RainFormer | 0.1782 | 0.2925 | 0.1107 | 0.1916 |
| CNN1D | 0.2049 | 0.3505 | 0.1307 | 0.2393 |
| Vanilla Transformer | 0.2985 | 0.2929 | 0.2092 | 0.2066 |

### D. Bounded-scope consistency check

We apply the synthetic-trained models directly to the measured *BIT-25* set, which is never used for training and contributes no calibration to the pipeline. The evaluation is therefore a zero-shot transfer test in the strict sense [47]. It asks whether physics-constrained synthetic training yields physically meaningful estimates on real data, not whether it attains operational accuracy.

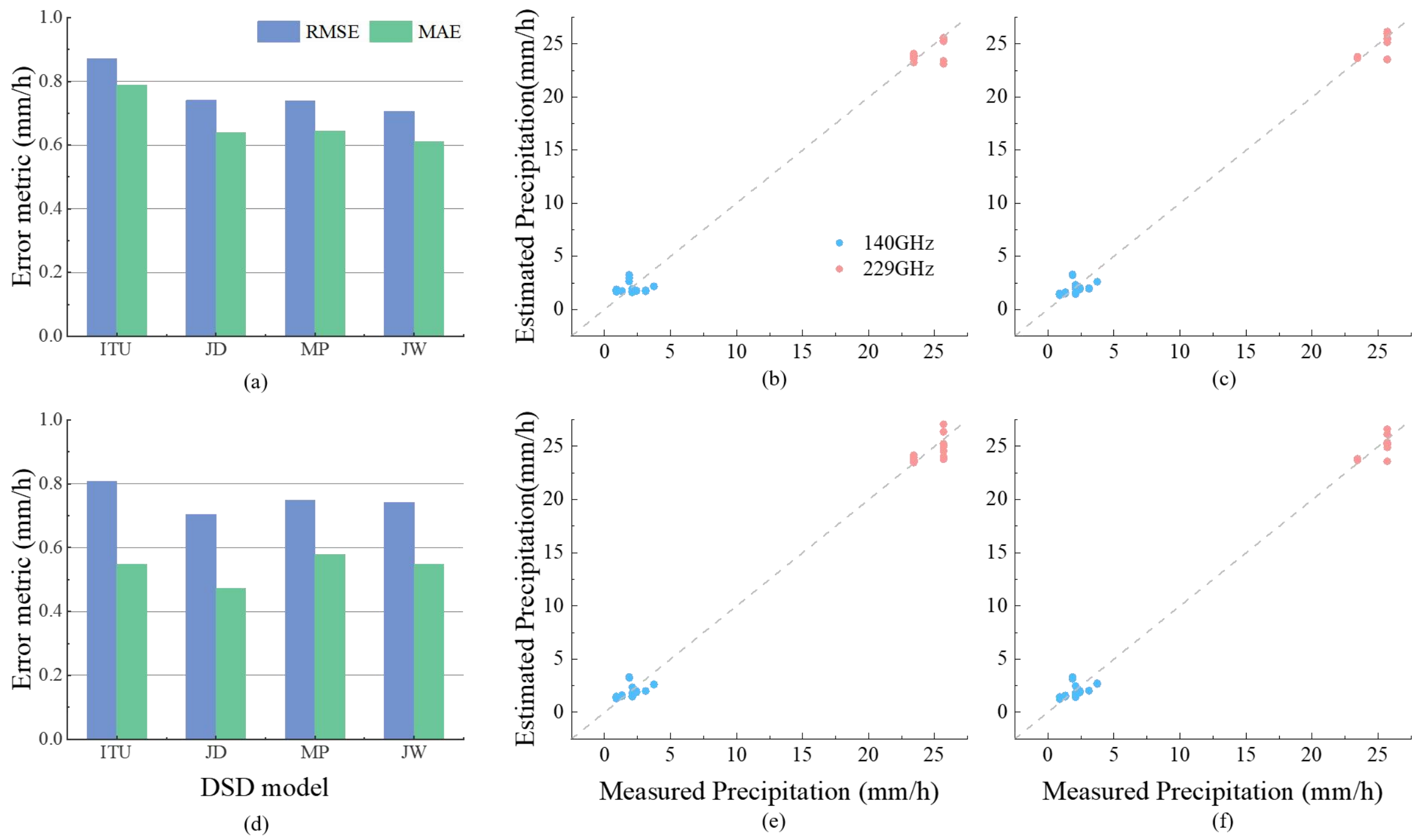

**Fig. 4** Measured-data consistency evaluation of RainFormer trained under different DSD priors. Panels (a) and (d) show the RMSE and MAE under different DSD priors at 140 and 229 GHz, respectively. Panels (b), (c), (e), and (f) show the corresponding estimated-versus-measured rainfall-rate scatter plots for the ITU, J-D, M-P, and J-W priors, respectively. The dashed line denotes the ideal 1:1 relationship.

Fig. 4 summarizes the error statistics and estimation behavior. Transfer error depends systematically on the DSD prior at both frequencies - direct evidence that the estimates are governed by the attenuation-rainfall physics embedded in training rather than by incidental dataset structure. Under every prior, the estimates preserve a monotonic relationship with gauge observations and cleanly separate weak from heavier rain, while the four priors jointly bracket the measured response, converting the unobservable path DSD from an unquantified bias into a bounded uncertainty envelope. The residual deviation from the 1:1 line reflects the expected domain gap: path-nonuniform rainfall, the mismatch between point-gauge labels and path-integrated attenuation - a known representativeness limit of opportunistic-link sensing [25] - background fluctuations, instrumental uncertainty, and the difference between true and assumed DSD.

The prior dependence differs across frequencies. At 140 GHz the empirical priors yield lower or more balanced errors than ITU-R, indicating the measured event is better represented by empirical microphysics under the present calibration. At 229 GHz the inter-prior spread narrows, though the small sample count there limits this to a qualitative observation. Consistent with the synthetic regression of Section V-A, no prior is universally preferred, reinforcing the bracketing strategy. The central result stands. Models trained purely on physics constrained synthetic data, without any measured-rainfall calibration, produce rank-consistent and physically interpretable estimates on an independent measured set.

### *E. Comparison With Physics-Based Retrieval*

We benchmark the learned estimator against two physics baselines that require no additional measurements. The first baseline inverts the ITU-R P.838-3 power law $\gamma = kR^{\alpha}$, giving the per-window rain rate $\hat{R}=(A/kL)^{1/\alpha}$, where $A$ is the mean rain-induced path attenuation, $L$=41.5 m, and $k$, $\alpha$ are the vertically polarized coefficients interpolated to each carrier, namely (1.559, 0.651) at 140 GHz and (1.639, 0.632) at 229 GHz. The second inverts, for each drop-size-distribution prior, the same Mie attenuation-rainrate relation used to generate the synthetic corpus, so that the physics baseline and the learned estimator share identical prior information. All baselines are evaluated on the measured *BIT-25* windows and scored by RMSE and MAE against the collocated gauge, with $5^{th}$-95th percentile bootstrap intervals at 140 GHz.

As shown in Table V, across all DSD priors the learned estimator reduces the measured-data RMSE far below the physics-only inversions. At 140 GHz it achieves 0.69 against 1.10 for the P.838-3 inversion and 1.98-4.52 for the Mie inversions, and at 229 GHz it achieves 0.69 against 4.07 for the P.838-3 inversion and 3.12-29.21 for the Mie inversions. Because only $N$=6 windows at a single gauge rate are available at 229 GHz, we present the 229 GHz comparison qualitatively. The 229 GHz figures should be read as illustrative of the direction and approximate magnitude of the improvement rather than as validated accuracy. Two regimes explain the 140 GHz behaviour. The window-mean attenuation there is only 0.06-0.19 dB, comparable to the 0.043 dB power-meter uncertainty, and the inversion nonlinearity $1/\alpha = 1.5$ amplifies this noise, so direct power-law inversion systematically overestimates weak rain. At 229 GHz the J-D inversion is the strongest physics baseline, consistent with the measured samples lying close to the J-D curve. The cross-prior spread of the physics inversions - from 3.12 to 29.21 mm/h at 229 GHz - quantifies the drop-size-distribution ambiguity, whereas the learned estimator collapses this spread to 0.69, effectively removing the DSD degree of freedom. We stress, however, that at 140 GHz the P.838-3 baseline's $5^{th}$-$95^{th}$ percentile bootstrap interval overlaps the learned estimator's point estimates. The 140 GHz improvement should therefore be read as a consistent trend rather than a statistically resolved gain, while the larger 229 GHz margins remain qualitative given $N$=6.

TABLE V

RAINFALL RETRIEVAL ON MEASURED *BIT-25* (RMSE/MAE).

| Method | RMSE | | MAE | |
|---|---|---|---|---|
| | 140GHz | 229GHz | 140GHz | 229GHz |
| P.838-3 inversion | 1.10 | 4.07 | 0.67 | 3.14 |
| Mie inversion (J-D) | 1.98 | 3.12 | 1.42 | 2.85 |
| Mie inversion (M-P) | 3.37 | 18.96 | 2.58 | 18.12 |
| Mie inversion (J-W) | 4.52 | 29.21 | 3.62 | 28.39 |
| RainFormer (best prior) | 0.69 | 0.69 | 0.60 | 0.47 |

## VI. CONCLUSION

This article presented a measurement-informed framework for opportunistic rainfall estimation using lower-THz received-power sequences. Outdoor measurements over a 41.5-m rain-exposed path at 140 and 229 GHz were used to characterize rain-induced attenuation and the rainfall dependence of an effective Rician K-factor. These propagation characteristics, together with ITU-R P.838-3 and Mie-theory attenuation relations under several canonical DSD assumptions, were incorporated into a physics-constrained simulator for training RainFormer. Under matched synthetic conditions, RainFormer achieved RMSEs of 0.1782 mm/h at 140 GHz and 0.2925 mm/h at 229 GHz. The ablation results show that temporal-order information and explicit attenuation-fluctuation statistics provide the largest contributions, whereas self-attention acts primarily as a refinement. RainFormer also achieved the most consistent performance across the investigated frequencies and metrics when compared with the 1D-CNN and vanilla Transformer baselines. Its application to independent measured data at 140 GHz produced rainfall estimates that were physically consistent with the gauge observations, indicating that short-term channel fluctuations can supplement mean attenuation when the rain-induced path loss is weak.

The results also demonstrate that rainfall retrieval from THz links is sensitive to the assumed propagation model and DSD, particularly when direct attenuation inversion is applied over a short path. The spread among the ITU-R and Mie-based scenarios should therefore be interpreted as model-form sensitivity rather than as a probabilistic uncertainty bound. The conclusions are limited by the single-site, single-event measurement campaign, the small measured dataset, the single-intensity 229-GHz subset, and the use of a scalar power receiver rather than a coherent channel sounder. Accordingly, the present results establish feasibility and physical consistency rather than operational retrieval accuracy. Broader validation will require multi-event and multi-site measurements, independent DSD observations, target-independent model selection, and evaluation across different path lengths, polarizations, frequencies, and rainfall regimes.

## REFERENCES

[1] B. Rong, "6G: The Next Horizon: From Connected People and Things to Connected Intelligence," IEEE Wireless Communications, vol. 28, no. 5, pp. 8-8, 2021.

[2] F. Liu et al., "Integrated Sensing and Communications: Toward Dual-Functional Wireless Networks for 6G and Beyond," IEEE Journal on Selected Areas in Communications, vol. 40, no. 6, pp. 1728-1767, 2022.

[3] W. Jiang et al., "Terahertz Communications and Sensing for 6G and Beyond: A Comprehensive Review," IEEE Communications Surveys & Tutorials, vol. 26, no. 4, pp. 2326-2381, 2024.

[4] A. M. Elbir, K. V. Mishra, S. Chatzinotas, and M. Bennis, "Terahertz-Band Integrated Sensing and Communications: Challenges and Opportunities," IEEE Aerospace and Electronic Systems Magazine, vol. 39, no. 12, pp. 38-49, 2024.

[5] F. Norouzian et al., "Rain Attenuation at Millimeter Wave and Low-THz Frequencies," IEEE Transactions on Antennas and Propagation, vol. 68, no. 1, pp. 421-431, 2020.

[6] Y. Song et al., "Terahertz channel power and BER performance in rain," Opt Express, vol. 33, no. 5, pp. 11336-11349, Mar 10 2025.

[7] A. Hirata et al., "Effect of Rain Attenuation for a 10-Gb/s 120-GHz-Band Millimeter-Wave Wireless Link," IEEE Transactions on Microwave Theory and Techniques, vol. 57, no. 12, pp. 3099-3105, 2009.

[8] J. Nemarich, R. J. Wellman, and J. Lacombe, "Backscatter and attenuation by falling snow and rain at 96, 140, and 225 GHz," IEEE Transactions on Geoscience and Remote Sensing, vol. 26, no. 3, pp. 319-329, 1988.

[9] L. Luini et al., "The Impact of Rain on Short E-Band Radio Links for 5G Mobile Systems," IEEE Trans. Antennas Propag., vol. 68, no. 4, pp. 3124 -3134, 2019.

[10] S. Ishii, M. Kinugawa, S. Wakiyama, S. Sayama, and T. Kamei, "Rain Attenuation in the Microwave-to-Terahertz Waveband," Wireless Engineering and Technology, vol. 07, no. 02, pp. 59-66, 2016.

[11] J. F. Federici, J. Ma, and L. Moeller, "Review of weather impact on outdoor terahertz wireless communication links," Nano Communication Networks, 10, pp. 12-26, 2016.

[12] P. Li et al., "Measurement and Modeling on Terahertz Channels in Rain," Journal of Infrared, Millimeter, and Terahertz Waves, vol. 46, no. 1, 2024.

[13] J. M. Riera et al., "Characterization of Rain Attenuation in 80–200 GHz From Experimental Drop Size Distributions," IEEE Trans. Antennas Propag., vol. 71, no. 5, pp. 4371-4379, 2023.

[14] G. Skofronick-Jackson et al., "The Global Precipitation Measurement (Gpm) Mission for Science and Society," Bull Am Meteorol Soc, vol. 98, no. 8, pp. 1679-1695, Aug 2017.

[15] G. J. Huffman et al., "Integrated Multi-satellite Retrievals for the Global Precipitation Measurement (GPM) Mission (IMERG)," in Satellite Precipitation Measurement, (Advances in Global Change Research, 2020, ch. Chapter 19, pp. 343-353.

[16] R. J. Joyce, J. E. Janowiak, P. A. Arkin, and P. Xie, "CMORPH: A Method that Produces Global Precipitation Estimates from Passive Microwave and Infrared Data at High Spatial and Temporal Resolution," Journal of Hydrometeorology, vol. 5, no. 3, pp. 487-503, 2004.

[17] T. Kubota et al., "Global Satellite Mapping of Precipitation (GSMaP) Products in the GPM Era," in Satellite Precipitation Measurement, (Advances in Global Change Research, 2020, ch. Chapter 20, pp. 355-373.
[18] Ashouri H et al., "PERSIANN-CDR: Daily Precipitation Climate Data Record from Multisatellite Observations for Hydrological and Climate Studies," Bulletin of the American Meteorological Society, vol. 96, no. 1, pp. 69-83, 2015.
[19] V. A. Gorooh, A. A. Asanjan, P. Nguyen, K. Hsu, and S. Sorooshian, "Deep Neural Network High Spatiotemporal Resolution Precipitation Estimation (Deep-STEP) Using Passive Microwave and Infrared Data," Journal of Hydrometeorology, vol. 23, no. 4, pp. 597-617, 2022.
[20] C. Wang, G. Tang, and P. Gentine, "PrecipGAN: Merging Microwave and Infrared Data for Satellite Precipitation Estimation Using Generative Adversarial Network," Geophysical Research Letters, vol. 48, no. 5, 2021.
[21] R. Rahimi, P. Ravirathinam, A. Ebtehaj, A. Behrangi, J. Tan, and V. Kumar, "Global Precipitation Nowcasting of Integrated Multi-satellitE Retrievals for GPM: A U-Net Convolutional LSTM Architecture," Journal of Hydrometeorology, vol. 25, no. 6, pp. 947-963, 2024.
[22] G. Ayzel, T. Scheffer, and M. Heistermann, "RainNet v1.0: a convolutional neural network for radar-based precipitation nowcasting," Geoscientific Model Development, vol. 13, no. 6, pp. 2631-2644, 2020.
[23] M. J. Piran, X. Wang, H. J. Kim, and H. H. Kwon, "Precipitation nowcasting using transformer-based generative models and transfer learning for improved disaster preparedness," International Journal of Applied Earth Observation and Geoinformation, vol. 132, 2024.
[24] H. Messer, A. Zinevich, and P. Alpert, "Environmental monitoring by wireless communication networks," Science, vol. 312, no. 5774, p. 713, 2006.
[25] H. Leijnse, R. Uijlenhoet, and J. N. M. Stricker, "Rainfall measurement using radio links from cellular communication networks," Water Resources Research, vol. 43, no. 3, 2007.
[26] K. Pu et al., "Rain Effects on Low-Frequency NLOS Microwave Links: Theoretical Modeling and Detectability Analysis," IEEE Trans. Antennas Propag., vol. 74, no. 4, pp. 3618-3632, 2026.
[27] J. Polz, C. Chwala, M. Graf, and H. Kunstmann, "Rain event detection in commercial microwave link attenuation data using convolutional neural networks," Atmospheric Measurement Techniques, vol. 13, no. 7, pp. 3835-3853, 2020.
[28] C. Chwala and H. Kunstmann, "Commercial microwave link networks for rainfall observation: Assessment of the current status and future challenges," WIREs Water, vol. 6, no. 2, 2019.
[29] B. Lian, Z. Wei, X. Sun, Z. Li, and J. Zhao, "A Review on Rainfall Measurement Based on Commercial Microwave Links in Wireless Cellular Networks," Sensors (Basel), vol. 22, no. 12, Jun 10 2022.
[30] A. Hirata, H. Takahashi, T. Kosugi, K. Murata, K. Naoya, and Y. Kado, "Rain Attenuation Statistics for a 120-GHz-Band Wireless Link," IEEE MTT-S International Microwave Symposium Digest, pp. 933, 2009.
[31] ITU-R, "Specific attenuation model for rain for use in prediction methods," Geneva, Switzerland, Recommendation ITU-R P.838-3, 2005.
[32] L. Yangang, "Statistical theory of the Marshall-Palmer distribution of raindrops," Atmospheric Environment. Part A. General Topics, vol. 27, no. 1, pp. 15-19, 1993.
[33] E. Feloni, K. Kotsifakis, N. Dervos, G. Giavis, and E. Baltas, Analysis of Joss-Waldvogel disdrometer measurements in rainfall events (Fifth International Conference on Remote Sensing and Geoinformation of the Environment (RSCy2017)). SPIE, 2017.
[34] C. Mätzler, "Drop-Size Distributions and Mie Computations for Rain," Institut für Angewandte Physik, Universität Bern, Bern, Switzerland, November 2002.
[35] R. F. Rincon and R. H. Lang, "Microwave link dual-wavelength measurements of path-average attenuation for the estimation of drop size distributions and rainfall," IEEE Transactions on Geoscience and Remote Sensing, vol. 40, no. 4, pp. 760-770, 2002.
[36] T. C. van Leth, H. Leijnse, A. Overeem, and R. Uijlenhoet, "Estimating raindrop size distributions using microwave link measurements: potential and limitations," Atmospheric Measurement Techniques, vol. 13, no. 4, pp. 1797-1815, 2020.
[37] X. Hao, T. S. Rappaport, R. J. Boyle, and J. H. Schaffner, "Measurements and models for 38-GHz point-to-multipoint radiowave propagation," IEEE Journal on Selected Areas in Communications, vol. 18, no. 3, pp. 310-321, 2000.
[38] B. Efron, "Bootstrap Methods: Another Look at the Jackknife," The Annals of Statistics, vol. 7, no. 1, 1979.
[39] L. Bernado, T. Zemen, F. Tufvesson, A. F. Molisch, and C. F. Mecklenbrauker, "Time- and Frequency-Varying K-Factor of Non-Stationary Vehicular Channels for Safety-Relevant Scenarios," IEEE Transactions on Intelligent Transportation Systems, vol. 16, no. 2, pp. 1007-1017, 2015.
[40] A. Abdi, C. Tepedelenlioglu, M. Kaveh, and G. Giannakis, "On the estimation of the K parameter for the Rice fading distribution," IEEE Communications Letters, vol. 5, no. 3, pp. 92-94, 2001.
[41] O. Peters and K. Christensen, "Rain: relaxations in the sky," Phys Rev E Stat Nonlin Soft Matter Phys, vol. 66, no. 3 Pt 2A, p. 036120, Sep 2002.
[42] J. Xu, et al. "Understanding and improving layer normalization." presented at the Advances in neural information processing systems 32, 2019.
[43] A. Vaswani et al., "Attention is All you Need," presented at the Advances in Neural Information Processing Systems 30, 2017.
[44] Y. N. Dauphin, A. Fan, M. Auli, and D. Grangier, "Language Modeling with Gated Convolutional Networks," presented at the Proceedings of the 34th International Conference on Machine Learning, 2017.
[45] N. Srivastava, G. Hinton, A. Krizhevsky, I. Sutskever, and R. Salakhutdinov, "Dropout: A Simple Way to Prevent Neural Networks from Overfitting," (in English), J Mach Learn Res, vol. 15, pp. 1929-1958, Jun 2014.
[46] H. Deng, G. Runger, E. Tuv, and M. Vladimir, "A time series forest for classification and feature extraction," Information Sciences, vol. 239, pp. 142-153, 2013.
[47] J. Tobin, R. Fong, A. Ray, J. Schneider, W. Zaremba, and P. Abbeel, "Domain randomization for transferring deep neural networks from simulation to the real world," presented at the 2017 IEEE/RSJ International Conference on Intelligent Robots and Systems (IROS), 2017.
[48] G. E. Karniadakis, I. G. Kevrekidis, L. Lu, P. Perdikaris, S. Wang, and L. Yang, "Physics-informed machine learning," Nature Reviews Physics, vol. 3, no. 6, pp. 422-440, 2021.
[49] A. Zeng, M. Chen, L. Zhang, and Q. Xu, "Are Transformers Effective for Time Series Forecasting?," Proceedings of the AAAI Conference on Artificial Intelligence, vol. 37, no. 9, pp. 11121-11128, 2023.
[50] Y. Zhang, S. Chen, W. Tian, and S. Chen, "Radar Reflectivity and Meteorological Factors Merging-Based Precipitation Estimation Neural Network," Earth and Space Science, vol. 8, no. 10, 2021.
[51] J. Pudashine, et al., "Deep learning for an improved prediction of rainfall retrievals from commercial microwave links." Water Resources Research 56, no. 7, pp. e2019WR026255, 2020.
[52] Q. Wen et al., "Transformers in Time Series: A Survey," presented at the Proceedings of the Thirty-Second International Joint Conference on Artificial Intelligence, 2023.
[53] Y. Nie, N. H. Nguyen, P. Sinthong, and J. Kalagnanam, "A Time Series Is Worth 64 Words: Long-term Forecasting with Transformers," presented at the International Conference on Learning Representations, 2023.
[54] T. Chai and R. R. Draxler, "Root mean square error (RMSE) or mean absolute error (MAE)? – Arguments against avoiding RMSE in the literature," Geoscientific Model Development, vol. 7, no. 3, pp. 1247-1250, 2014.
[55] A. Gulati et al., "Conformer: Convolution-augmented Transformer for Speech Recognition," presented at the Interspeech 2020, 2020.